\documentclass[showpacs, oneside, twocolumn, prd, amsmath, amssymb, nofootinbib, superscriptaddress]{revtex4-1}

\pdfoutput=1
\usepackage{cases}
\usepackage{amsmath}
\usepackage{amssymb}
\usepackage{amsfonts}
\usepackage{amssymb}
\usepackage{dcolumn}
\usepackage{bm}
\usepackage{bbm}
\usepackage{graphicx}
\usepackage{xcolor}
\usepackage{array}
\usepackage{subfigure}
\usepackage{hyperref}
\usepackage{wasysym}
\usepackage{mathrsfs}

\newcommand{\be}{\begin{equation}}
\newcommand{\ee}{\end{equation}}
\newcommand{\ba}{\begin{eqnarray}}
\newcommand{\ea}{\end{eqnarray}}

\newcommand{\gsim}{\mathrel{\hbox{\rlap{\lower.55ex \hbox {$\sim$}}
			\kern-.3em \raise.4ex \hbox{$>$}}}}
\newcommand{\lsim}{\mathrel{\hbox{\rlap{\lower.55ex \hbox {$\sim$}}
			\kern-.3em \raise.4ex \hbox{$<$}}}}

\hypersetup{colorlinks=true,
	breaklinks=true,
	pdfstartview=Fit,
	linkcolor=blue,
	citecolor=blue,
	urlcolor=blue}

\begin{document}

\title{Interpreting cosmological tensions from the effective field theory of torsional gravity}

\author{Sheng-Feng Yan}
\affiliation{Department of Astronomy, School of Physical Sciences, University of Science and Technology of China, Hefei, Anhui 230026, China}
\affiliation{CAS Key Laboratory for Researches in Galaxies and Cosmology, University of Science and Technology of China, Hefei, Anhui 230026, China}
\affiliation{School of Astronomy and Space Science, University of Science and Technology of China, Hefei, Anhui 230026, China}

\author{Pierre Zhang}
\affiliation{Department of Astronomy, School of Physical Sciences, University of Science and Technology of China, Hefei, Anhui 230026, China}
\affiliation{CAS Key Laboratory for Researches in Galaxies and Cosmology, University of Science and Technology of China, Hefei, Anhui 230026, China}
\affiliation{School of Astronomy and Space Science, University of Science and Technology of China, Hefei, Anhui 230026, China}

\author{Jie-Wen Chen}
\affiliation{Department of Astronomy, School of Physical Sciences, University of Science and Technology of China, Hefei, Anhui 230026, China}
\affiliation{CAS Key Laboratory for Researches in Galaxies and Cosmology, University of Science and Technology of China, Hefei, Anhui 230026, China}
\affiliation{School of Astronomy and Space Science, University of Science and Technology of China, Hefei, Anhui 230026, China}

\author{Xin-Zhe Zhang}
\affiliation{Department of Astronomy, School of Physical Sciences, University of Science and Technology of China, Hefei, Anhui 230026, China}
\affiliation{CAS Key Laboratory for Researches in Galaxies and Cosmology, University of Science and Technology of China, Hefei, Anhui 230026, China}
\affiliation{School of Astronomy and Space Science, University of Science and Technology of China, Hefei, Anhui 230026, China}

\author{Yi-Fu Cai}
\email{yifucai@ustc.edu.cn}
\affiliation{Department of Astronomy, School of Physical Sciences, University of Science and Technology of China, Hefei, Anhui 230026, China}
\affiliation{CAS Key Laboratory for Researches in Galaxies and Cosmology, University of Science and Technology of China, Hefei, Anhui 230026, China}
\affiliation{School of Astronomy and Space Science, University of Science and Technology of China, Hefei, Anhui 230026, China}

\author{Emmanuel N. Saridakis}
\email{msaridak@phys.uoa.gr}
\affiliation{Department of Physics, National Technical University of Athens, Zografou Campus GR 157 73, Athens, Greece}
\affiliation{National Observatory of Athens, Lofos Nymfon, 11852 Athens, Greece}
\affiliation{Department of Astronomy, School of Physical Sciences, University of Science and Technology of China, Hefei, Anhui 230026, China}

\begin{abstract}
Cosmological tensions can arise within $\Lambda$CDM scenario amongst different observational windows, which may indicate new physics beyond the standard paradigm if confirmed by measurements. In this article, we report how to alleviate both the $H_0$ and $\sigma_8$ tensions simultaneously within torsional gravity from the perspective of effective field theory (EFT). Following these observations, we construct concrete models of Lagrangians of torsional gravity. Specifically, we consider the parametrization $f(T)=-T-2\Lambda/M_P^2+\alpha T^\beta$, where two out of the three parameters are independent. This model can efficiently fit observations solving the two tensions. To our knowledge, this is the first time where a modified gravity theory can alleviate both $H_0$ and $\sigma_8$ tensions simultaneously, hence, offering an additional argument in favor of gravitational modification.
\end{abstract}



\maketitle

\section{Introduction}
As the standard paradigm, the Lambda cold dark matter ($\Lambda$CDM) cosmology has been tested by various observations, from which the acceleration of today's Universe is interpreted to be sourced from a cosmological constant. However, the nature of cosmic acceleration remains mysterious. The possibility of a dynamical dark energy (DE), as well as the need of a stochastic process such as inflation to generate initial conditions that seed the large-scale structures (LSS) led to many proposals based either on the introduction of new fields \cite{Peebles:2002gy, Cai:2009zp} or on gravity theories beyond general relativity (GR) \cite{Capozziello:2011et, Nojiri:2010wj}.

With the accumulation of cosmological data, experimental tensions may arise within $\Lambda$CDM cosmology. If they were to remain under the increasing precision of experimental observations, they would constitute, in a statistical sense, clear indications of new physics beyond $\Lambda$CDM. One recently well-debated tension relates the value of the Hubble parameter at present
time $H_0$ measured from the cosmic microwave background (CMB) temperature and polarization data by the Planck Collaboration to be $H_0 = 67.37 \pm 0.54 \ \mathrm{km \, s^{-1} \, Mpc^{-1}}$ \cite{Aghanim:2018eyx}, to the one from local measurements of the Hubble Space Telescope yielding $H_0 = 74.03 \pm 1.42 \ \mathrm{km \, s^{-1} \, Mpc^{-1}}$ \cite{Riess:2019cxk}. Recent analyses with the combination of gravitational lensing and time-delay effects data reported a significant deviation at $5.3 \sigma$ \cite{Wong:2019kwg}. Another potential tension concerns the measurements of the parameter $\sigma_8$, which
quantifies the gravitational clustering of matter from the amplitude of the linearly evolved power spectrum at the scale of $8 h^{-1} \text{Mpc}$. Specifically, a possible deviation was noticed between measurements of CMB and LSS surveys, namely, between Planck \cite{Aghanim:2018eyx} and SDSS/BOSS \cite{Zarrouk:2018vwy, Alam:2016hwk, Ata:2017dya}. Nevertheless, the statistical
confidence of this cosmological ``tension'' remains low and is not as manifest as the $H_0$ tension \cite{Bohringer:2014ooa, Bhattacharyya:2018fwb, Kazantzidis:2018rnb}. Although these two tensions could in principle arise from unknown systematics, the
possibility of  physical origin puts the standard lore of cosmology into additional investigations, by pointing to various extensions beyond $\Lambda$CDM.

In this article, we consider systematically the $H_0$ and $\sigma_8$ tensions, and report how to alleviate both simultaneously within torsional gravity. We exploit the effective field theory (EFT) of torsional gravity, a formalism that allows for a systematic investigation of the background and perturbations separately. This approach was developed early on in \cite{ArkaniHamed:2003uy} for curvature gravity, and recently in \cite{Li:2018ixg, Cai:2018rzd} for torsional gravity. EFT approaches have been widely applied with success in cosmology, for instance, to inflation \cite{Cheung:2007st, ArkaniHamed:2007ky} and DE \cite{Gubitosi:2012hu, Bloomfield:2012ff, Gleyzes:2013ooa}.
In order to address cosmological tensions, we identify the effects of gravitational modifications within the EFT on the dynamics of the background and of linear perturbation levels. This will allow us to construct specific models of $f(T)$ gravity providing adequate deviation from $\Lambda$CDM that can alleviate $H_0$ and $\sigma_8$ tensions.

\section{Theoretical background}
%
The $H_0$ tension can be alleviated in specific theoretical models, such as early DE \cite{Poulin:2018cxd}, interacting DE \cite{Yang:2018qmz}, dark radiation \cite{Archidiacono:2013fha}, 
improved big bang nucleosynthesis (BBN) \cite{Bernal:2016gxb},
or modified gravity \cite{Nunes:2018xbm, Cai:2019bdh}. Also, the $\sigma_8$ tension may be addressed by sterile neutrinos \cite{Guo:2018ans}, running vacuum models \cite{Gomez-Valent:2018nib,Bloomfield:2012ff}, a dark matter (DM) sector that clusters differently at small and large scales \cite{Kunz:2015oqa}, or by modified gravity \cite{Kazantzidis:2018rnb}. There were attempts from th nonconventional matter sector to address both tensions simultaneously, such as DM-neutrino interactions \cite{DiValentino:2017oaw}.
%
However, it is remarkable to question whether both tensions can be alleviated simultaneously via gravitational modifications in a much general framework. Indeed, the $H_0$ tension reveals a universe that is expanding faster at late times than that from a cosmological constant preferred by CMB data, while a lower value of $\sigma_8$ than the one of CMB most likely $\Lambda$CDM would imply that matter clusters either later on or less efficiently. Hence, these two observations seem to indicate that there might be ``less gravitational power'' at intermediate scales, which phenomenologically advocates a possible modification of gravitation. Accordingly, in this article we address the aforementioned question within the EFT framework for torsional gravity.


For a general curvature-based gravity, the action following the EFT approach in the unitary gauge, invariant by space diffeomorphisms, which expanded around a flat FRW metric $ds^2= -dt^2+a^2(t)\,\delta_{ij} dx^i dx^j$, is given by
\begin{align}
\label{curvEFTact}
S &= \int d^4x \Big\{ \sqrt{-g} \big[ \frac{M^2_P}{2} \Psi(t)R - \Lambda(t) - b(t)g^{00}
\nonumber
\\
& +M_2^4(\delta g^{00})^2 -\bar{m}^3_1 \delta g^{00} \delta K -\bar{M}^2_2 \delta K^2 -\bar{M}^2_3
\delta K^{\nu}_{\mu} \delta K^{\mu}_{\nu} \nonumber \\
& + m^2_2 h^{\mu\nu}\partial_{\mu} g^{00}\partial_{\nu}g^{00} +\lambda_1\delta R^2
+\lambda_2\delta R_{\mu\nu}\delta R^{\mu\nu} +\mu^2_1 \delta g^{00} \delta R \big]
\nonumber \\
& +\gamma_1 C^{\mu\nu\rho\sigma} C_{\mu\nu\rho\sigma} +\gamma_2
\epsilon^{\mu\nu\rho\sigma} C_{\mu\nu}^{\quad\kappa\lambda} C_{\rho\sigma\kappa\lambda}
\nonumber \\
& +\sqrt{-g} \big[ \frac{M^4_3}{3}(\delta g^{00})^3 -\bar{m}^3_2(\delta g^{00})^2 \delta K + ... \big] \Big\} ~,
\end{align}
where $M_P = (8\pi G_N)^{-{1}/{2}}$ is the reduced Planck mass with $G_N$ the Newtonian constant. $R$ is the Ricci scalar corresponding to the Levi-Civit$\grave{\mathrm{a}}$ connection, $C^{\mu\nu\rho\sigma}$ is the Weyl tensor, $\delta K^{\nu}_{\mu}$ is the perturbation of the extrinsic curvature, and the functions $\Psi(t)$, $\Lambda(t)$, $b(t)$ are determined by the background evolution.

When the underlying theory also includes torsion \cite{Li:2018ixg}, one can generalize the EFT action as follows,
\begin{align}
\label{actionfin}
S = & \int d^4x \sqrt{-g} \Big[ \frac{M^2_P}{2} \Psi(t)R - \Lambda(t) - b(t) g^{00} +
\frac{M^2_P}{2} d(t) T^0 \Big] \nonumber \\
& + S^{(2)} ~.
\end{align}
To compare with the effective action for curvature-based gravity \eqref{curvEFTact}, one reads that at background level there is additionally the zeroth part $T^0$ of the contracted torsion tensor $T^{0\mu}_{\ \mu}$, with its time-dependent coefficient $d(t)$.
Furthermore, the perturbation part $S^{(2)}$ contains all operators of the perturbation part of \eqref{curvEFTact}, plus pure torsion terms including $\delta T^2$, $\delta T^0\delta T^0$, and $\delta T^{\rho\mu\nu}\delta T_{\rho\mu\nu}$, and extra terms that mix curvature and torsion, namely, $\delta T\delta R$, $\delta g^{00}\delta T$, $\delta g^{00}\delta T^0$, and $\delta K\delta T^0$, where $T\equiv\frac{1}{4} T^{\rho \mu \nu} T_{\rho \mu \nu} + \frac{1}{2} T^{\rho \mu \nu
} T_{\nu \mu\rho } - T_{\rho \mu }^{\ \ \rho }T_{\ \ \ \nu}^{\nu \mu }$ is the torsion scalar.
Adding the matter action $S_m$ to the effective action of torsional-based gravity (\ref{actionfin}) and then performing variation, one obtains the Friedmann equations to be \cite{Li:2018ixg}:
\begin{align}
\label{f11}
H^2 &= \frac{1}{3 M_{P}^2} \big( \rho_m +\rho_{DE}^{\text{eff}} \big) ~, \\
\dot{H} &= -\frac{1}{2 M_{P}^2} \big( \rho_m +\rho_{DE}^{\text{eff}} +p_m +p_{DE}^{\text{eff}} \big) ~, \nonumber
\end{align}
and where
\begin{align}
\label{rhopDE}
\rho_{DE}^{\text{eff}} &= b+\Lambda -3 M_P^2 \Big[ H\dot{\Psi}+\frac{dH}{2}+H^2(\Psi-1) \Big] ~, \\
p_{DE}^{\text{eff}} &= b -\Lambda +M_P^2 \Big[ \ddot{\Psi} +2H\dot{\Psi} +\frac{\dot{d}}{2} +(H^2 +2\dot{H})(\Psi -1) \Big] ~, \nonumber
\end{align}
are, respectively, the effective DE density and pressure in the general torsional gravity. Moreover, we treat the matter sector as dust that satisfies the conservation equation $\dot{\rho}_m +3H(\rho_m +p_m) =0$, which in terms of redshift leads to $\rho_m = 3 M_P^2 H_0^2 \Omega_{m0}(1+z)^3$, with $\Omega_{m0}$ the value of $\Omega_{m}\equiv8\pi G_N\rho_m/(3H^2)$ at present.

\section{Model independent analyses}
%
In general, to avoid the $H_0$ tension one needs a positive correction to the first Friedmann equation at late times that could yield an increase in $H_0$ compared to the $\Lambda$CDM scenario. As for the $\sigma_8$ tension, we recall that in any cosmological model, at sub-Hubble scales and through the matter epoch, the equation that governs the evolution of matter perturbations in the
linear regime is \cite{Lue:2004rj, Linder:2005in, Uzan:2006mf, Gannouji:2008wt, Anagnostopoulos:2019miu}
\begin{eqnarray}
\label{eq:delta-evolution}
\ddot{\delta}+2 H \dot{\delta} = 4 \pi G_{\mathrm{eff}} \rho_m \delta ~,
\end{eqnarray}
where $G_{\mathrm{eff}}$ is the effective gravitational coupling given by a generalized Poisson equation [see, e.g., \cite{Gleyzes:2013ooa} for an explicit expression of $G_{\mathrm{eff}}$ for the operators present in the action~\eqref{curvEFTact}]. In general, $G_{\mathrm{eff}}$ differs from the Newtonian constant $G_{N}$, and thus contains information from gravitational modifications (note that $G_{\mathrm{eff}}=G_{N}$ in $\Lambda$CDM cosmology). Solving for $\delta(a)$ provides the
observable quantity $f\sigma_8(a)$, following the definitions $f(a)\equiv d\ln \delta(a)/d\ln a$ and $\sigma(a) = \sigma_8
\delta(1)/\delta(a=1)$. Hence, alleviation of the $\sigma_8$ tension may be obtained if $G_{\mathrm{eff}}$ becomes
smaller than $G_{N}$ during the growth of matter perturbations and/or if the ``friction'' term in
\eqref{eq:delta-evolution} increases.

To grasp the physical picture, we start with a simple case: $b(t) = 0$ and $\Lambda(t)=\Lambda=const$ [$b$ and $\Lambda$ are highly degenerate as shown in \eqref{rhopDE}], while $\Psi(t)=1$. Hence, from \eqref{actionfin}, with the above coefficient choices, the only deviation from $\Lambda$CDM at the background level comes from the term $d(t)T^{0}$, and we remind that in FRW geometry
$T^{0}=H$ when evaluated on the background. In this case, the first Friedmann equation in \eqref{f11},  using for convenience the redshift $z=-1+a_0/a$ as the dimensionless variable and setting $a_0=1$, yields
\begin{eqnarray}
\label{eq:H}
H(z) = -\frac{d(z)}{4} +\sqrt{\frac{d^2(z)}{16} +H_{\Lambda \text{CDM}}^2(z)} ~,
\end{eqnarray}
where $H_{\Lambda \text{CDM}}(z) \equiv H_0 \sqrt{\Omega_m(1+z)^3+\Omega_\Lambda}$ is the Hubble rate in $\Lambda$CDM, with $\Omega_m=\rho_m/(3M_p^2H^2)$ the matter density parameter and primes denote derivatives with respect to $z$.
Accordingly, if $d<0$ and is suitably chosen, one can have $H(z\rightarrow z_{\rm CMB}) \approx H_{\Lambda\text{CDM}}(z\rightarrow z_{\rm CMB})$ but $H(z\rightarrow 0) > H_{\Lambda\text{CDM}}(z\rightarrow 0)$; i.e., the $H_0$ tension is solved [one should choose $|d(z)| < H(z)$, and thus, since $H(z)$ decreases for smaller $z$, the deviation from $\Lambda$CDM will be significant only at low redshift]. Additionally, since the friction term in \eqref{eq:delta-evolution} increases, the growth of structure gets damped, and therefore, the $\sigma_8$ tension is also solved [note that since we have imposed $\Psi=1$, then $G_{\mathrm{eff}}=G_N$ as one can verify from \eqref{actionfin} and \eqref{f11}; namely, the contributions from $T^0$ vanish at first order in perturbations].

Furthermore, for typical values that lie well within (or to the closest of) the $1\sigma$ intervals of the $H(z)$ redshift surveys, it is expected that CMB measurements will be sensitive to such a deviation from the $\Lambda$CDM scenario for nonvanishing $T^0$ at early times. Actually, the $T^0$ operator acts in a similar way as a conventional cosmological constant. Thus, it adds yet another new functional form to parametrize the background and leads to more flexibility in fitting redshift and clustering measurements. Because of the fact that $\Lambda$ and $T^0$ are highly degenerate, an interesting possibility is to ask whether a universe without a cosmological constant but with a boundary term containing $T^0$ can fit the data well. To assess such a possibility, a fully consistent numerical analysis including both CMB and redshift measurements is required. This gives interesting consequences for various probes in the intermediate-to-high redshift range accessible to ongoing and near-future target surveys such as quasars, Lyman-$\alpha$, or 21-cm lines.

\section{$f(T)$ gravity}
%
Next, we propose concrete models of torsional modified gravity that can be applied to alleviate the two cosmological tensions based on the torsional EFT dictionary. In particular, we focus on the well-known class of torsional gravity, namely, the $f(T)$ gravity \cite{Cai:2015emx}, which is characterized by the action $S=\frac{M_P^2}{2} \int d^4x e f(T)$, with $e = \text{det}(e_{\mu}^A) = \sqrt{-g}$ and $e^A_\mu$ the vierbein, and thus by the Friedmann equations $H^2= \frac{\rho_m}{3M_P^2} +\frac{T}{6} -\frac{f}{6} +\frac{Tf_T}{3}$, $\dot{H} = -\frac{1}{2 M_{P}^2} (\rho_m+p_m) +\dot{H}(1+f_{T}+2Tf_{TT})$, with $f_{T}\equiv\partial f/\partial T$, $f_{TT} \equiv \partial^{2} f/\partial T^{2}$, where we have applied $T=6H^2$ in flat FRW geometry (we follow the convention of \cite{Li:2018ixg}). Therefore, $f(T)$ gravity can arise from the general EFT approach to torsional gravity by choosing $\Psi=-f_T$, $\Lambda=\frac{M_P^2}{2}(T f_T-f)$, $b=0$, $d=2\dot {f}_T$ \cite{Li:2018ixg}, and can restore GR by choosing $f(T)=-T-2\Lambda/M_P^2$. Similarly, one can use the EFT approach to describe $f(R,T)$ gravity, and hence, $f(T,B)$ gravity too, where $B=-2\nabla_{\mu} T_\nu^{\ \nu\mu}$ is the boundary term in the relation $R=-T+B$.

The above EFT approach holds for every $f(T)$ gravity by making a suitable identification of the involved time-dependent functions. For instance, we consider the following ansatz:
 $f(T)=-[T+6H_0^2(1-\Omega_{m0})+F(T)] $,
where $F(T)$ describes the deviation from GR [note, however, that in FRW geometry, apart from the regular choice $F=0$, the $\Lambda$CDM scenario can also be obtained for the special case $F(T)=c ~ T^{1/2}$ too, with $c$ a constant]. Under these assumptions, the first Friedmann equation becomes
\begin{align}
\label{eq:bg3}
 T(z)+2\frac{F'(z)}{T'(z)} T(z)-F(z)= 6H^2_{\Lambda CDM}(z) ~.
\end{align}
In order to solve the $H_0$ tension, we need $T(0) = 6 H_0^2 \simeq 6 (H_0^{CC})^{2}$, with $H_0^{CC}=74.03$ $ \mathrm{km \, s^{-1} \, Mpc^{-1}}$ following the local measurements \cite{Riess:2019cxk}, while in the early era of $z\gtrsim 1100$ we require the Universe expansion to evolve as
in $\Lambda$CDM, namely $H(z\gtrsim 1100) \simeq H_{\Lambda  CDM}(z\gtrsim 1100)$
\footnote{As mentioned earlier, this requirement follows from the fact that we are considering modifications kicking in only at late times, and therefore, the results in tension from CMB analysis performed within $\Lambda CDM$ remain unaffected. In particular, the expansion history being modified only at low redshift, the diameter distance proportional to the integral of the Hubble rate over time from now all the way to recombination is hardly unaffected. }.
This implies $F(z)|_{z\gtrsim 1100}\simeq c T^{1/2}(z)$ (the value $c=0$ corresponds to standard GR, while for $c\neq0$ we obtain $\Lambda$CDM too). Note that, in this case the effective gravitational coupling is given by
\cite{Nesseris:2013jea}
\begin{eqnarray}
\label{Geff}
 G_{\mathrm{eff}}=\frac{G_{N}}{1+F_{T}} ~.
\end{eqnarray}
Therefore, the perturbation equation at linear order \eqref{eq:delta-evolution} becomes
\begin{equation}
\label{eq:delta-z}
 \delta'' + \left[ \frac{T'(z)}{2T(z)} -\frac{1}{1+z} \right] \delta' = \frac{9 H_0^2 \Omega_{m0} (1+z) }{[1+F'(z)/T'(z)] T(z)} \delta ~,
\end{equation}
where $\delta\equiv\delta\rho_{m}/\rho_m$ is the matter overdensity. Since around the last scattering moment $z\gtrsim 1100$ the Universe should be matter-dominated, we impose $\delta'(z)|_{z\gtrsim 1100} \simeq -\frac{1}{1+z}\delta(z)$, while at late times we look for $\delta(z)$ that leads to an $f\sigma_8$ in agreement with redshift survey observations.

\begin{figure*}[ht]
\centering
\subfigure{\includegraphics[width=3.in]{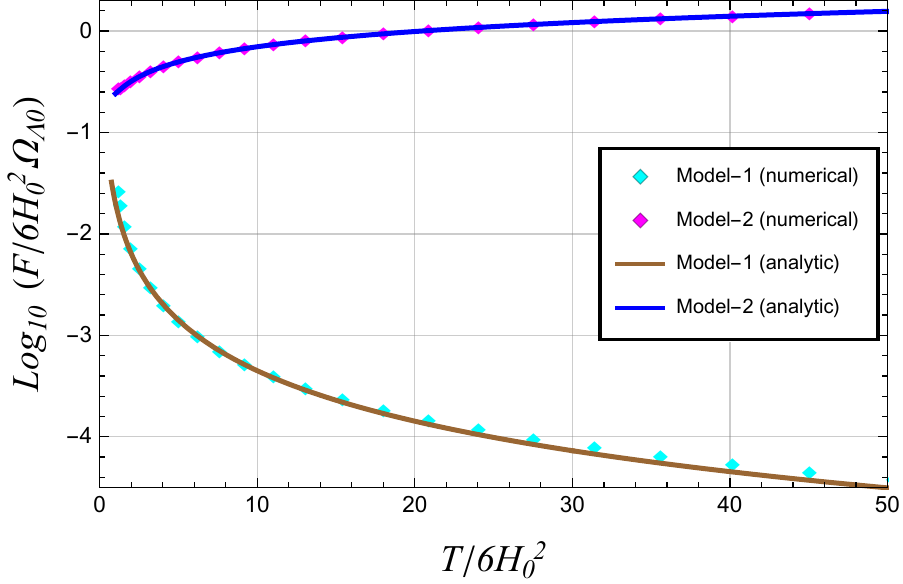}}
\subfigure{\includegraphics[width=3.05in]{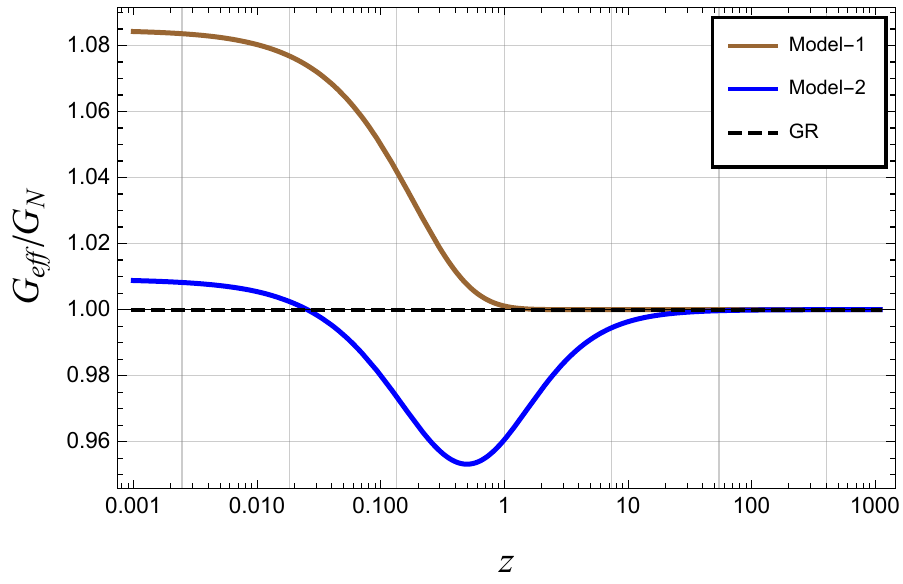}}
\caption{Left panel: Reconstruction of two $f(T)$ models. The cyan and magenta rhombic points denote the numerical results of Model-1 and Model-2, respectively; the brown and blue solid curves are the parametrizations given by Eqs.~\eqref{mod1} and \eqref{mod2}, respectively. Right panel: Redshift evolution of $G_{\mathrm{eff}}/G_N$ in Model-1 (brown solid line) and Model-2 (blue solid line) and their comparison to the GR case (black dashed line).}
\label{fig:fTmodels}
\end{figure*}

By solving \eqref{eq:bg3} and \eqref{eq:delta-z} with initial and boundary conditions at $z \sim 0$ and $z \sim 1100$, we can find the functional forms for the free functions of the $f(T)$ gravity that we consider, namely, $T(z)$ and $F(z)$, that can alleviate both $H_0$ and $\sigma_8$ tensions. In the left panel of Fig.~\ref{fig:fTmodels}, we depict two such forms for $F(T)$. Both models
approach the $\Lambda$CDM scenario at $z\gtrsim 1100$, with Model-1 approaching $F=0$ and hence restoring GR, while Model-2
approaches $F\propto T^{1/2}$, and thus it reproduces $\Lambda$CDM. In particular, we find that we can well fit the numerical solutions of Model-1 by
\begin{eqnarray}
\label{mod1}
 F(T) \approx 375.47 \Big( \frac{T}{6 H_0^2} \Big)^{-1.65} ~,
\end{eqnarray}
and of Model-2 by
\begin{equation}
\label{mod2}
 F(T) \approx 375.47 \Big( \frac{T}{6 H_0^2} \Big)^{-1.65} + 25 T^{1/2} ~.
\end{equation}
Note that, the first term of Model-2, which coincides with Model-1, provides a small deviation to $\Lambda$CDM at late times, while it decreases rapidly to become negligible in the early Universe. In addition, we examine $G_{\mathrm{eff}}$ given by \eqref{Geff} for the two models \eqref{mod1} and \eqref{mod2}, which are displayed in the right panel of Fig.~\ref{fig:fTmodels}. As expected, at high redshifts in both models, $G_{\mathrm{eff}}$ becomes $G_N$, recovering the $\Lambda$CDM paradigm. At very low redshifts, $G_{\mathrm{eff}}$ becomes slightly higher than $G_N$, increasing slightly the gravitational strength. This gravitational modification is in competition at late times with the accelerating expansion. It turns out that the effect of an increased cosmic acceleration with
respect to $\Lambda$CDM in our $f(T)$ gravity models dominates over the stronger gravitational strength in the clustering of matter. We check that both models can easily pass the BBN constraints (which demand $|G_{\mathrm{eff}}/G_N-1|\leq0.2$ \cite{Copi:2003xd}), as well as the ones from the Solar System [which demand $|G_{\mathrm{eff}}'(z=0)/G_N|\leq10^{-3}h^{-1}$ and $| G_{\mathrm{eff}}''(z = 0) / G_N | \leq 10^{5} h^{-2}$ \cite{Nesseris:2006hp}].

\begin{figure*}[ht]
\centering
\subfigure{\includegraphics[width=3.in]{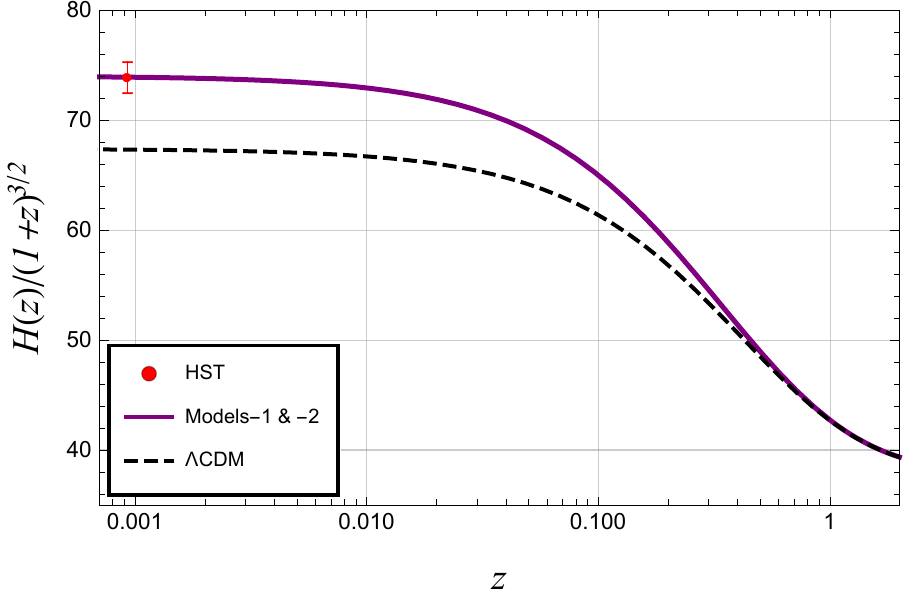}}
\subfigure{\includegraphics[width=3.1in]{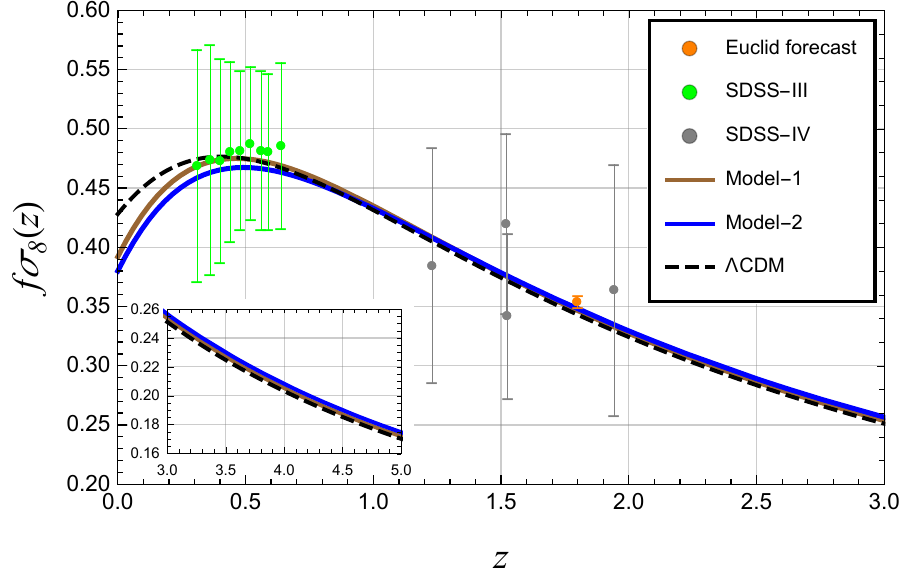}}
\caption{Left panel: Evolution of the Hubble parameter $H(z)$ in the two $f(T)$ models (purple solid line) and in $\Lambda$CDM cosmology (black dashed line). The red point represents the latest data from extragalactic Cepheid-based local measurement of $H_0$ provided in \cite{Riess:2019cxk}. Right panel: Evolution of $f\sigma_8$ in Model-1 (brown solid line) and Model-2 (blue solid line)
of $f(T)$ gravity and in $\Lambda$CDM cosmology (black dashed line). The green data points are from BAO observations in SDSS-III DR12 \cite{Wang:2017wia}, the gray data points at higher redshift are from SDSS-IV DR14 \cite{Gil-Marin:2018cgo,Hou:2018yny,Zhao:2018jxv}, while the red point around $\sim 1.8$ is the forecast from Euclid \cite{Taddei:2016iku}. The subgraph in the left bottom displays $f \sigma_8$ at high redshift $z = 3 \sim 5$, which shows that the curve of Model-2 is above the one of Model-1 and $\Lambda$CDM scenario and hence approaches $\Lambda$CDM slower than Model-1.}
\label{fig:H0&fs8}
\end{figure*}

\begin{figure*}[ht]
\centering
\subfigure{\includegraphics[width=3.in]{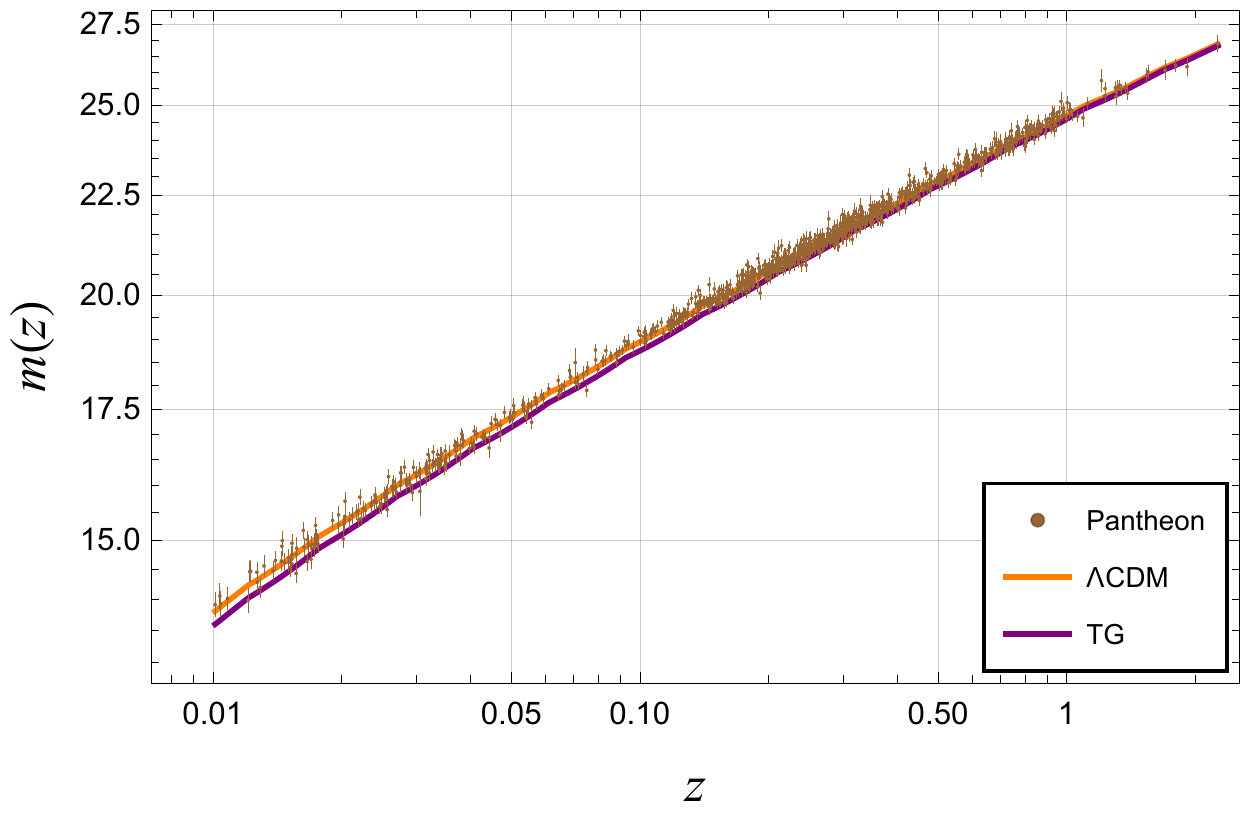}}
\subfigure{\includegraphics[width=3.in]{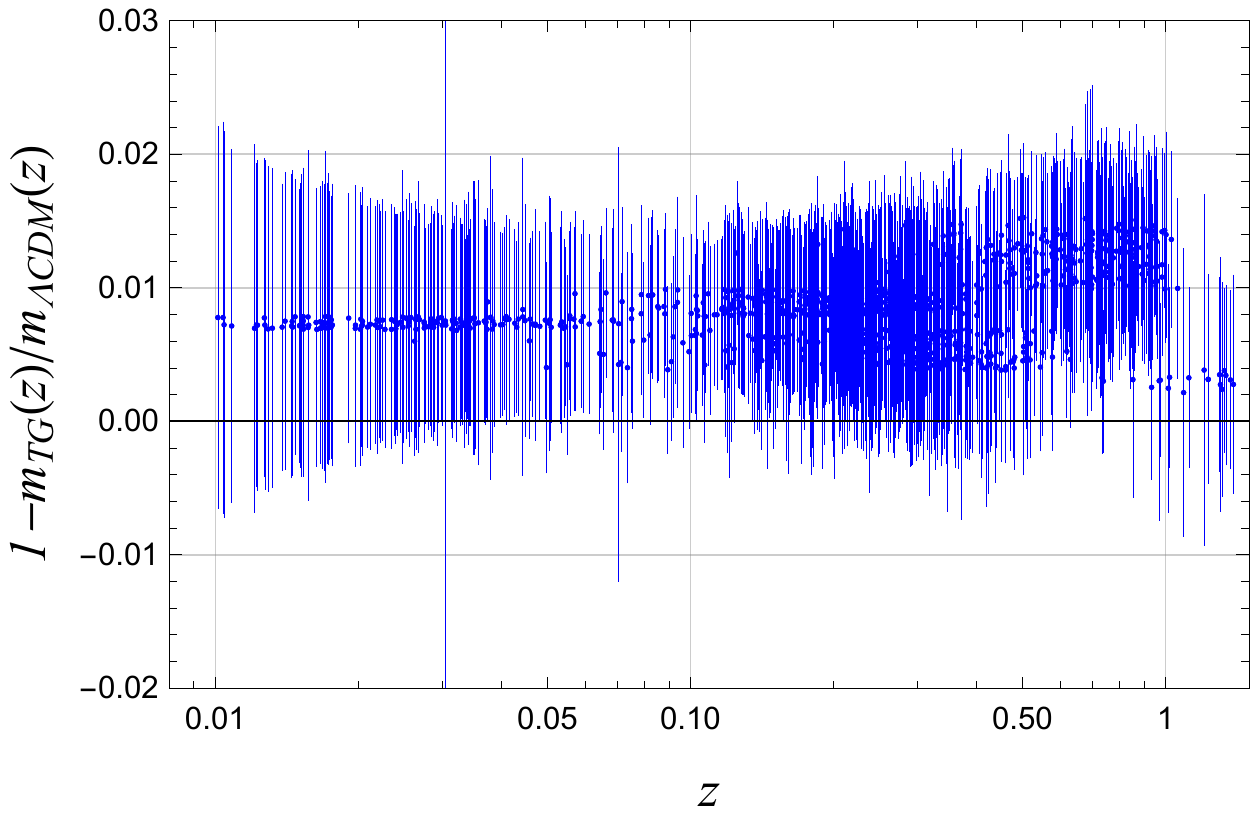}}
\caption{Left panel: Distance modulus magnitude $m = 5 \mathrm{log}_{10} D_L(z)+25+M$ in TG and $\Lambda$CDM with Planck close to best fit $H_0$ and $M = -19.45$ (Pantheon close to best fit) vs Pantheon SN data. Right panel: Ratio of modulus distances in TG and $\Lambda$CDM vs Pantheon SN error bars divided by the data.}
\label{fig:Pantheon}
\end{figure*}

Now we show how Model-1 and Model-2 can alleviate the $H_0$ and $\sigma_8$ tension by solving the background and perturbation equations. In Fig.~\ref{fig:H0&fs8}, we present the evolution of $H(z)$ and $f\sigma_8$ for two $f(T)$ models, and we compare them with $\Lambda$CDM. We stress that the $H_0$ tension can be alleviated as $H(z)$ remains statistically consistent for all CMB and CC measurements at all redshifts. We remind the reader that the two $f(T)$ models differing merely by a term $\propto T^{1/2}$, which does not affect the background as explained before, are degenerate at the background level. However, at the perturbation level, the two models behave differently as their gravitational coupling $G_{\mathrm{eff}}$ differs. We further stress that both models can alleviate the $\sigma_8$ tension, and fit efficiently to BAO and LSS measurements. Note that at high redshifts ($z\geq2$), Model-2 approaches $\Lambda$CDM slower than Model-1, but in a way that is statistically indistinguishable for present-day data. Nevertheless, future high-redshift surveys such as eBOSS for quasars and Euclid \cite{Laureijs:2011gra} for galaxies have the potential to discriminate among the predictions of $f(T)$ gravity and the $\Lambda$CDM scenario. Moreover, the clusters and CMB measurements on $\sigma_8$ are in good agreement in our models, as the CMB preferred values in $\Lambda$CDM get further lowered than local ones from rescaling $\sigma_8$ by the ratio of the growth factors in $f(T)$ gravity and $\Lambda$CDM
\footnote{Explicitly,
\begin{equation}
\sigma_8^{f(T)}(z=0) = \frac{D^{f(T)}(z=0)}{D^{\Lambda}(z=0)} \frac{D^{\Lambda}(z_{\rm eff})}{D^{f(T)}(z_{\rm eff})} \sigma_8^{\Lambda}(z=0) ,
\end{equation}
where $D(z)$ is the growth factor, $f(T)$ and $\Lambda$ denote our models and $\Lambda$CDM respectively, and $z_{\rm eff}$ is the effective redshift of the measurements ($z_{\rm eff} \sim 0.1$ for clusters experiments and $z_{\rm eff} \sim 1100$ for CMB temperature fluctuations observations). It turns out that, as at high redshift, $z_{\rm eff} \sim 1100$, the growth factor is the same in either our $f(T)$-models or in $\Lambda CDM$, but at low redshift, $z_{\rm eff} \sim 0.1$, the growth factor is approximately 1.03 bigger in the later compared to the formers, cluster $\sigma_8$-measurements get bigger by about such an amount reducing the gap with CMB preferred value in this modified gravity scenario.}.
Moreover, in Fig.~\ref{fig:Pantheon} we illustrate how this scheme is valid against to the Pantheon catalog \cite{Benevento:2020fev}. According to the plots, the difference between Model-1 (labeled as TG in plots) and $\Lambda$CDM is well within the error bars, as well as the residuals are consistent with zero. Note that in a real fit to Pantheon, there are even more room with free varying $M$ and $\Omega_m$: the residuals between Model-1 and data will be smaller.

In short summary, we conclude that the class of $f(T)$ gravity: $f(T) = -T -2\Lambda/M_P^2 +\alpha T^\beta$, where only two out of the three parameters $\Lambda$, $\alpha$, and $\beta$ are independent (the third one is eliminated using $\Omega_{m0}$), can alleviate both $H_0$ and $\sigma_8$ tensions with suitable parameter choices. Moreover, such kinds of models in $f(T)$ gravity could also be examined through galaxy-galaxy lensing effects \cite{Chen:2019ftv}, strong lensing effects around black holes \cite{Li:2019lsm} and gravitational wave experiments \cite{Cai:2018rzd}.

\section{Extensions in $f(T, B)$ gravity}
%
It is straightforward to generalize the EFT analysis into other torsional modifications that can also address the observational tensions. One such extension is $f(T, B)$ gravity, in which the Lagrangian is a function of both the torsion scalar $T$ and the boundary term $B=-2\nabla_{\mu} T_\nu^{\ \nu\mu} $ \cite{Bahamonde:2015zma} (note that in FRW geometry $B=6\dot{H}+18H^2$, with dots denoting derivatives with respect to cosmic time $t$). Here we consider the subclass $f(T) = -T + F(B)$. In this case, the correspondence with the EFT parameters is: $\Psi(t) = 1$, $b(t) = 0$, $d(t)= 2 \partial^2 F(B)/(\partial B\partial t)$. By fixing $d(t) = const$, we acquire that $F_B$ evolves linearly with cosmic time $t$. One can then solve the evolution equations for $d(t)$ and $\Lambda (t)$, imposing the observational measurements, in order to reconstruct the form of $f(T, B)$, as was done for $f(T)$ gravity. Extension to more general cases of gravity, where $\Psi (t)$, $b(t)$, as well as higher-order operators, are left as free functions, can be considered along the lines developed here. We leave concrete model building in $f(T, B)$ gravity and other modified gravity theories for future work.

\section{Conclusion and discussions}
%
In this article, we reported how theories of torsional gravity can alleviate both $H_0$ and $\sigma_8$ tensions simultaneously. Working within the EFT framework, torsional gravity theories can be identified as the EFT operators that allow us to extract the evolution equations of the background and of the perturbations in a model-independent manner. This allows us to address in a systematic way how tensions amongst the observational measurements, such as the ones on $H_0$ and $\sigma_8$, can be relaxed. Following these considerations, we constructed concrete models from specific Lagrangians, describing cosmological scenarios where these tensions fade away. In particular, we investigated the well-known $f(T)$ gravity. Imposing initial conditions at the last scattering that reproduce the $\Lambda$CDM scenario, and imposing the late-time values preferred by local measurements, we reconstructed two particular forms of $f(T)$. These models are well described by the parametrization: $f(T) = -T -2 \Lambda/M_P^2 +\alpha T^\beta$. To our knowledge, this is the first time where both $H_0$ and $\sigma_8$ tensions are simultaneously alleviated by a modified gravity theory.

We mention that we used the simplest approach of EFT to torsional gravity, in the sense that we considered only operators present at the background level. Although we found parametrizations of the lowest-order operators efficient in alleviating cosmological tensions, constructing more sophisticated scenarios by involving extra operators can lead to a fruitful phenomenology and thus inspire further investigations in many different directions. Namely, it would be interesting to perform an observational confrontation using various datasets to assess in a statistical way the performance of the modified gravity theory that we considered. Moreover, one can use the information from the EFT to construct gravitational modifications beyond the $f(T)$ class that can solve both tensions simultaneously, such as the $f(T, B)$ extensions, symmetric teleparallel gravity, $f(T, T_G)$ gravity, etc. These topics, while interesting and necessary, lie beyond the scope of this first investigation, and shall be addressed in follow-up works.

To end, we point out that our results can be put in the perspective of forthcoming LSS surveys that cover intermediate-to-high redshift ranges, such as probes of quasars, Lyman-$\alpha$, or emission lines, where not only higher values of $H(z)$ could be detected, but also lower values of $f\sigma_8$ following the suppression of the structure growth from the early start of the
accelerated expansion. These surveys will shed light and help probe the observable effects predicted by torsional gravity.

\section*{Acknowledgments}
%
We are grateful to F. Anagnostopoulos, S. Basilakos, D. Easson, C. Li, X. Ren, L. Senatore, and H. Wong for discussions.
Y. F. C is supported in part by the NSFC (Grants No. 11722327, No.  11653002, No. 11961131007, No. 11421303), by the CAST-YESS (Grant No. 2016QNRC001), and by the Fundamental Research Funds for Central Universities.
E. N. S. is supported partly by the USTC fellowship for international visiting professors.
This article is partially based upon work from COST Action ``Cosmology and Astrophysics Network for Theoretical Advances and Training Actions'', supported by the European Cooperation in Science and Technology.
All numerics were operated on the computer clusters {\it LINDA} \& {\it JUDY} in the particle cosmology group at USTC.


\end{document}